\documentclass[12pt]{article}
\usepackage[a4paper, total={7in, 10in}]{geometry}
\usepackage[parfill]{parskip}
\usepackage{physics, tensor, float, subcaption}
\usepackage{graphicx}
\graphicspath{ {Plots/} }
\usepackage{jhep-mod}
\usepackage{bm}
\usepackage{soul}
\usepackage{amssymb,amsmath,amsthm}
\usepackage{mathrsfs}
\usepackage[utf8]{inputenc}
\usepackage{enumerate}
\usepackage{bigints}
\usepackage{xcolor}
\usepackage{appendix}
\usepackage{graphicx}
\usepackage{float}
\usepackage{tikz}
\usepackage{setspace}
\usepackage{cancel}
\usepackage{array}
\usepackage{tabulary}
\usepackage{doi}
\definecolor{purple}{rgb}{1,0,1}
\definecolor{lime}{HTML}{A6CE39} 

\definecolor{lime}{HTML}{A6CE39}
\newcommand{\orcidicon}{%
	\begin{tikzpicture}
	\draw[lime, fill=lime] (0,0) 
		circle [radius=0.16] 
		node[white] {{\fontfamily{qag}\selectfont \tiny ID}};
	\draw[white, fill=white] (-0.0625,0.095) 
		circle [radius=0.007];
	\end{tikzpicture}
	\hspace{-5mm}
}
\newcommand\orcidJosh{{\href{https://orcid.org/0000-0003-1200-7261}{\orcidicon}}}
\newcommand\orcidMatt{{\href{https://orcid.org/0000-0003-1088-6485}{\orcidicon}}}

\renewcommand{\O}{\mathcal{O}}

\newcommand{\be}{\begin{equation}gin{equation}}
\newcommand{\ee}{\end{equation}}

\def\A{{\mathcal{A}}}
\begin{document}
\newcommand{\arXiv}[1]{arXiv:\href{https://arxiv.org/abs/#1}{\color{blue}#1}}


\title{\vspace{-25pt}\huge{
Photon escape cones, \\
physical and optical metrics, 
asymptotic and near-horizon physics.\\
}}


\author{
\Large
Joshua Baines\!\orcidJosh {\sf  and} Matt Visser\!\orcidMatt}
\affiliation{School of Mathematics and Statistics, Victoria University of Wellington, 
\\
\null\qquad PO Box 600, Wellington 6140, New Zealand.}
\emailAdd{joshua.baines@sms.vuw.ac.nz}
\emailAdd{matt.visser@sms.vuw.ac.nz}

\abstract{
\vspace{1em}

We consider the explicit analytic behaviour of photon escape cones in generic static spherically symmetric spacetimes, emphasizing the interplay between the physical spacetime metric and the optical metric, and the interplay between 
large-distance asymptotic and near-horizon physics.
The circular photon orbits (photon spheres) are shown to be given by wormhole throats in the optical metric, (not the physical metric), and the escape cone solid angle is easily calculable in terms of the capture cross section $\sigma_{capture}$, the area of the spherical 2-surfaces, and the norm of the timelike Killing vector. 
Under appropriate circumstances, for near-horizon photon emission the escape cone solid angle can be related to the surface gravity $\kappa_H$.  We provide a number of illustrative examples, involving both black holes and wormholes, including situations with multiple photon spheres. 

\bigskip
\noindent
{\sc Date:} 26 August 2023; \LaTeX-ed \today

black hole; wormhole; black bounce. \\

\bigskip
\noindent{\sc PhySH:} 
Gravitation
}

\maketitle
\def\tr{{\mathrm{tr}}}
\def\diag{{\mathrm{diag}}}
\def\cof{{\mathrm{cof}}}
\def\pdet{{\mathrm{pdet}}}
\def\QED{ {\hfill$\Box$\hspace{-25pt}  }}
\def\d{{\mathrm{d}}}
\def\sign{\hbox{sign}}

\parindent0pt
\parskip7pt

\clearpage
\null
\vspace{-75pt}
\section{Introduction}

The study of escape cones for photons isotropically emitted from various locations in spacetime has a long history, dating back almost 60 years to early work by Synge in the mid 1960's~\cite{Synge:1966}. Synge was able to show that, for photons emitted from some radius $r_*$ in the standard Hilbert form of the Schwarzschild spacetime,\footnote{\;\;$\d s^2 = -(1-2m/r)\d t^2 + {\d r^2\over 1-2m/r} +r^2 \d \Omega^2$.} only those emitted within an escape cone, $\theta< \theta_*$, close to the local vertical could actually escape to spatial infinity. 

\begin{figure}[htbp]
\begin{center}
{%
\setlength{\fboxsep}{0pt}%
\setlength{\fboxrule}{2pt}%
\fbox{\includegraphics[scale=0.25]{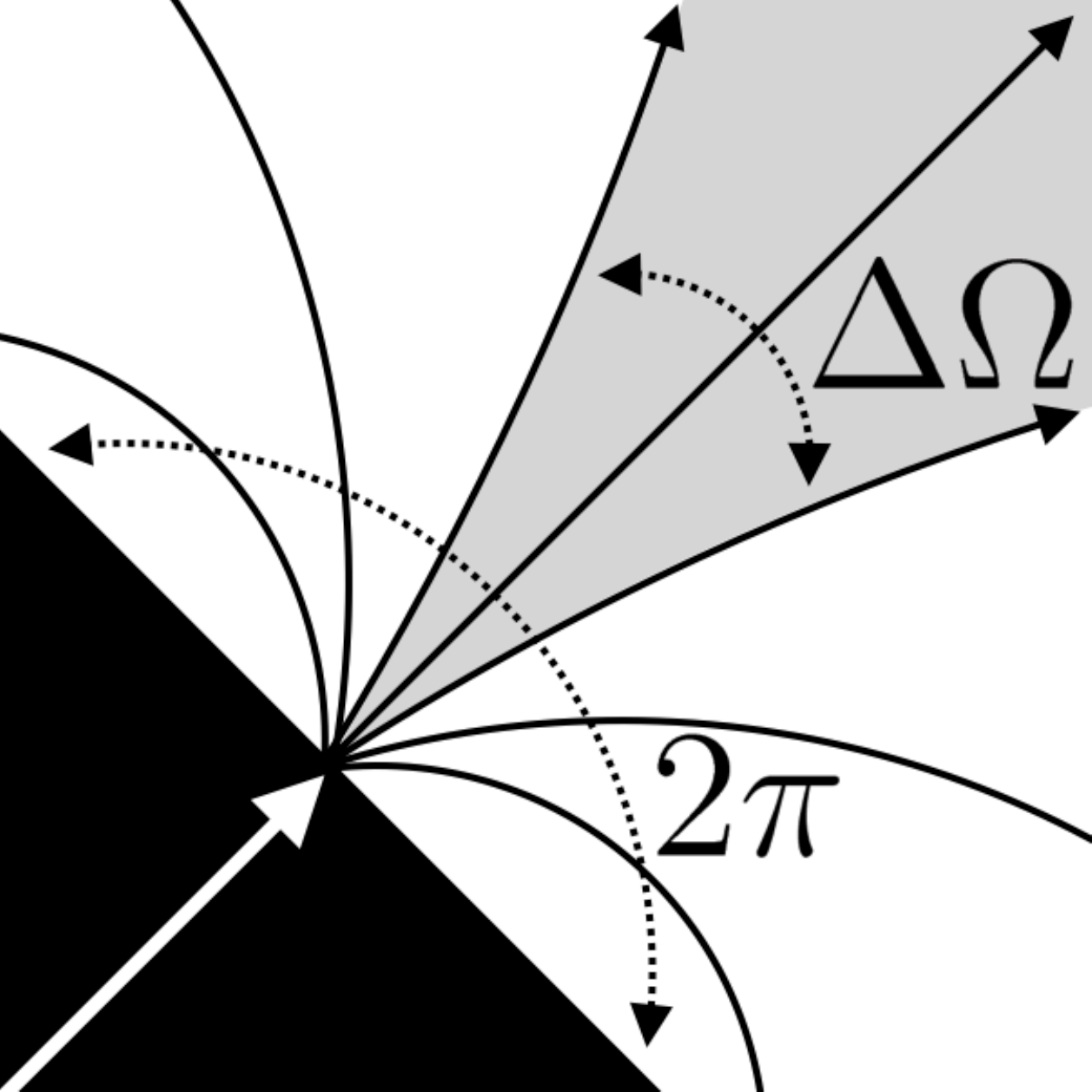}}
}%
\caption{Qualitative escape cone for photons emitted ar $r_*$.}
\label{F:cone}
\end{center}
\end{figure}

Specifically, Synge showed the equivalent of~\cite{Synge:1966}
\begin{equation}
\sin \theta_* = \sqrt{27} \;\; {m\over r_*} \;\sqrt{1-{2m\over r_*}},
\end{equation}
whence
\begin{equation}
\cos \theta_* =  -\left(1-{3m\over r_*}\right)  \;\sqrt{1+{6m\over r_*}}.
\end{equation}

If, instead of  considering the opening angle of the escape cone, one works with the solid angle subtended by the escape cone then (see for instance~\cite{phenomenology})
\begin{equation}\Delta\Omega(r_*)= 
\int_0^{\theta_*}  \sin\theta\; d\theta \int_0^{2\pi} d\phi= 2\pi(1-\cos\theta_*).
\end{equation}
Thence for Schwarzschild spacetime (see for instance~\cite{phenomenology})
\begin{equation}\Delta\Omega(r_*)
= 2\pi\left[ 1+ \left(1-{3m\over r_*}\right) \sqrt{1+{6m\over r_*}} \; \right].
\end{equation}\enlargethispage{20pt}
The reason for preferring to work with solid angle instead of opening angle is purely pragmatic: $P_*= {1\over4\pi} \;\Delta\Omega(r_*)$ has the direct physical interpretation of being  the escape probability of an isotropically emitted photon. 
Quantitatively understanding the escape probability $P_*$ is of central importance when analyzing near-horizon energy fluxes~\cite{energy-balance,SgA-energy-balance} and assessing the physical observability of horizons~\cite{observability}.

\begin{figure}[htbp]
\begin{center}
{%
\setlength{\fboxsep}{0pt}%
\setlength{\fboxrule}{1pt}%
\fbox{\includegraphics[scale=0.25]{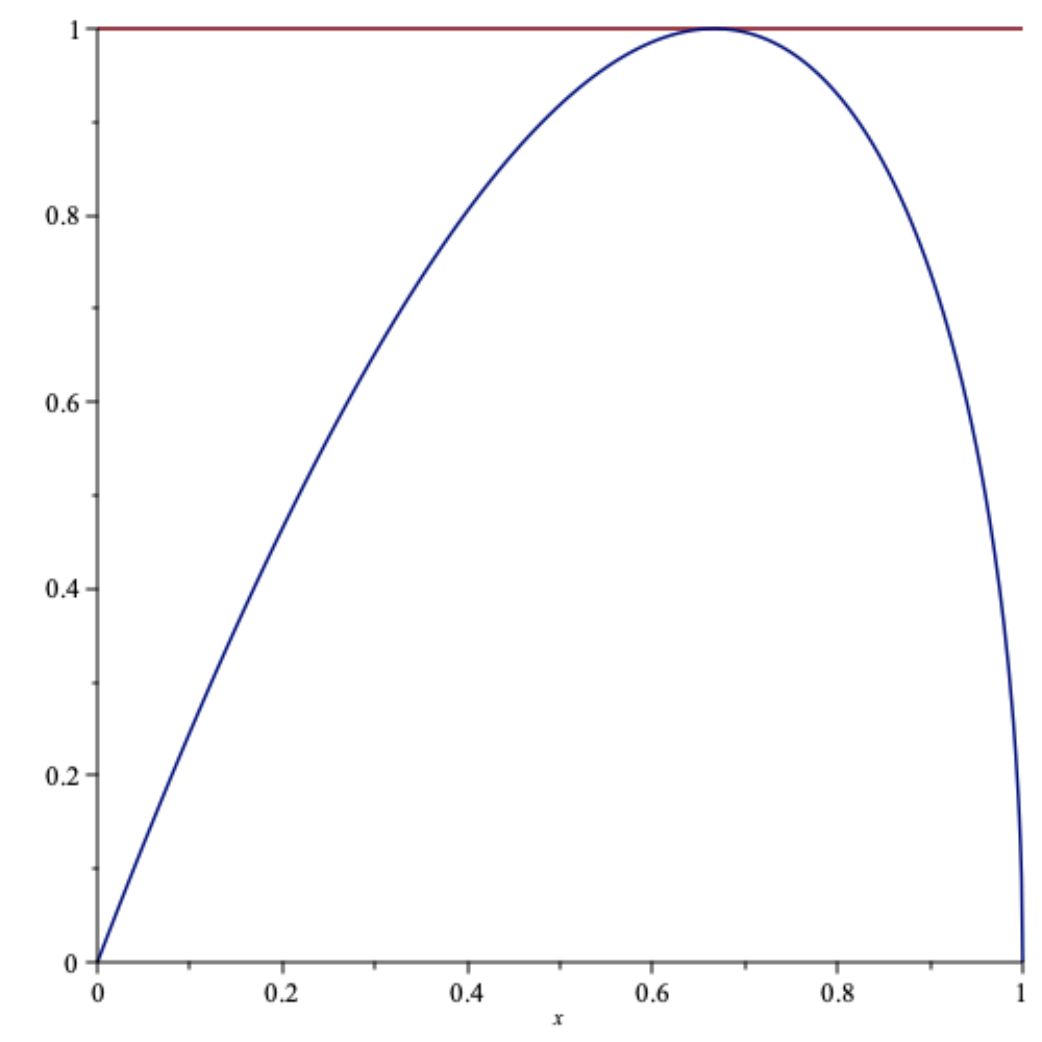}}
}%
\caption{Synge's $\sin\theta_*$ as a function of $2m/r_*$.}
\label{F:cone}
\end{center}
\end{figure}

\begin{figure}[htbp]
\begin{center}
{%
\setlength{\fboxsep}{0pt}%
\setlength{\fboxrule}{1pt}%
\fbox{\includegraphics[scale=0.25]{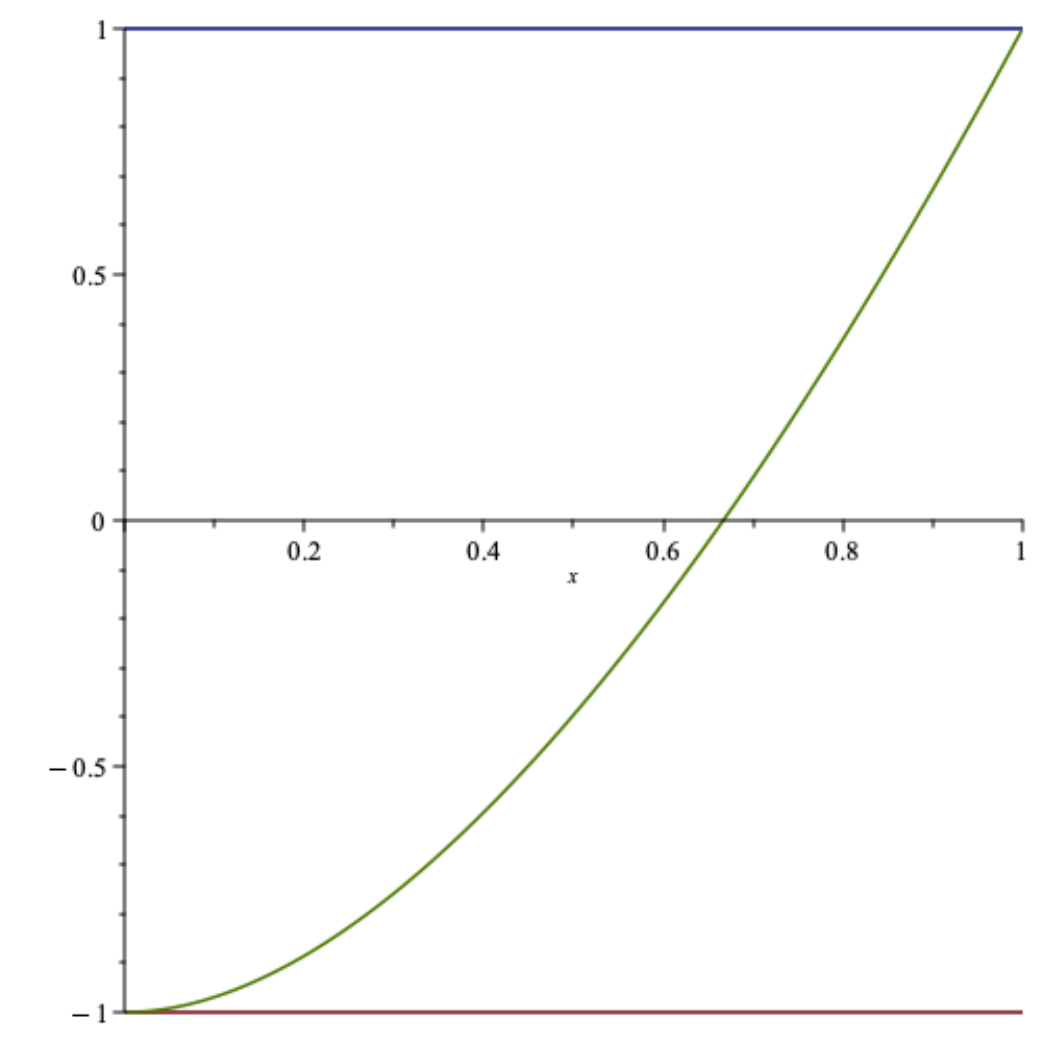}}
}%
\caption{Synge's (implicit) $\cos\theta_*$ as a function of $2m/r_*$.}
\label{F:cone}
\end{center}
\end{figure}

\begin{figure}[htbp]
\begin{center}
{%
\setlength{\fboxsep}{0pt}%
\setlength{\fboxrule}{1pt}%
\fbox{\includegraphics[scale=0.25]{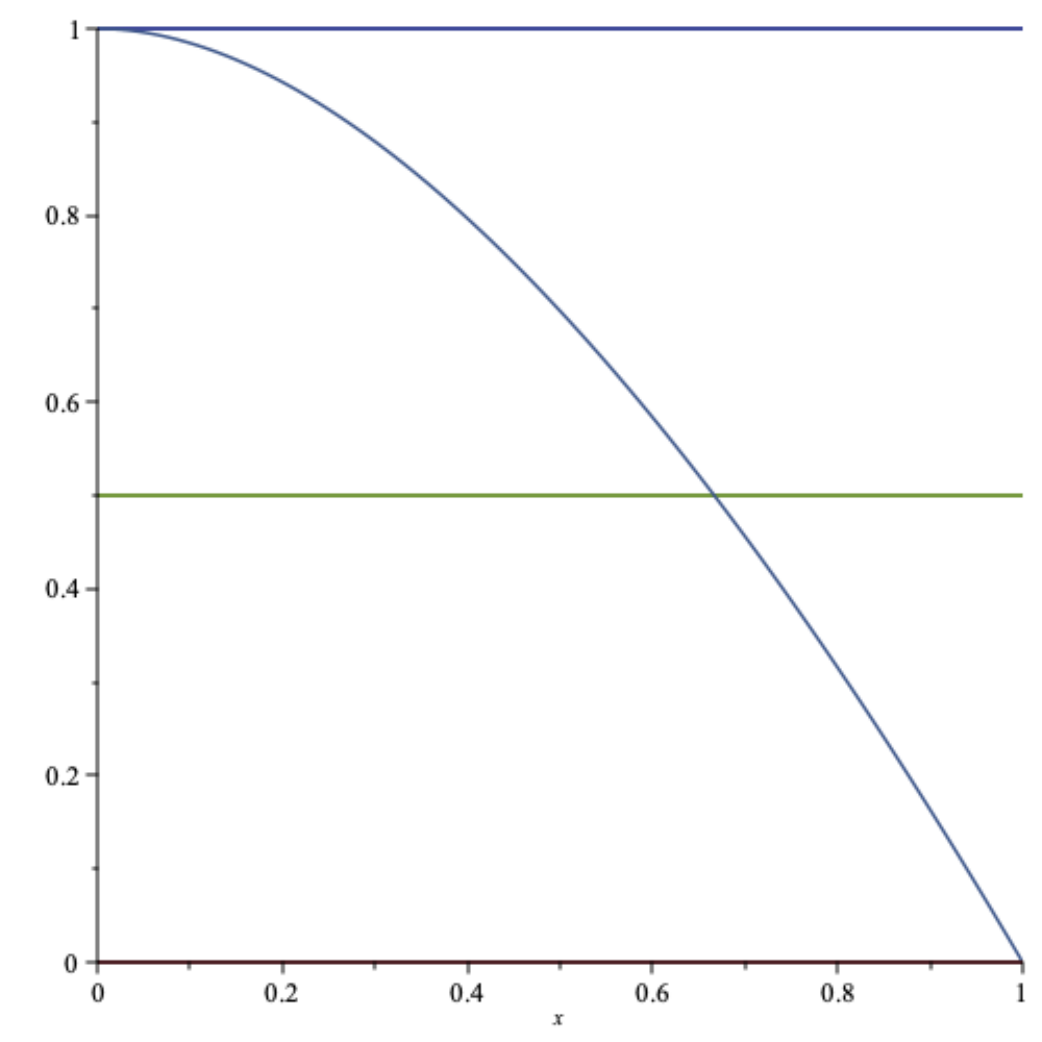}}
}%
\caption{Synge's (implicit) escape probability $P_*={\Delta\Omega_*\over4\pi}$ as a function of $2m/r_*$.}
\label{F:cone}
\end{center}
\end{figure}

Over the last 60 years considerable effort has been put into the study of escape cones, in Schwarzschild, Kerr, and various other phenomenologically inspired model spacetimes. The complement of the escape cone is the capture cone, $\Delta\Omega_{capture} = 4\pi -\Delta\Omega_*$, which is intimately related to the silhouette (often called the ``shadow'') of the object in question.  See for instance the recent review articles~\cite{Perlick:2021,Perlick:2010}, the analyses in~\cite{Tsupko:2017, Grenzebach:2014}, and the recent observational results in references~\cite{ETH-M87,ETH-SgA}.  Quite a lot can be said for generic static spherically symmetric spacetimes --- but herein we shall see that there are still a few interesting nuggets of information that can be extracted.  Wherever possible, we emphasize analytic insight over numerics.

\section{The physical metric}

Adopting (with only minor modifications) the notation of reference~\cite{Perlick:2021}, any static spherically symmetric spacetime can 
(without loss of generality) be cast into the manifestly static form
\begin{equation}\d s^2 = - A(r)\, \d t^2 + B(r)\, \d r^2 + D(r) \{\d\theta^2+\sin^2\theta\;\d\phi^2\}.
\end{equation}
This has the advantage that one is not \emph{a priori} enforcing unnecessary co-ordinate choices.
Now define
\begin{equation}H(r) = {D(r)\over A(r)}.
\end{equation}
--- A brief calculation (see~\cite{Perlick:2021}) shows that the location of the photon spheres $r_\gamma$ (circular photon orbits) are defined by solutions of
\begin{equation}r_\gamma: \quad {d H(r)\over dr}=0.
\end{equation}
--- In typical situations there will only be one (unstable) photon sphere between the outermost horizon and spatial infinity, or only one (unstable) photon sphere between the two asymptotic regions of a traversable wormhole, corresponding to a global minimum of $H(r)$. We put aside for now the complications due to possible multiple photon spheres. \\
--- A brief calculation (see~\cite{Perlick:2021}) shows that the opening angle $\theta_*$ of the escape cone (for light emitted from $r_*$ and escaping out to spatial infinity) is given by
\begin{equation}
\sin \theta_* =\sqrt {H(r_\gamma)\over H(r_*)}.
\end{equation}
Thence
\begin{equation}
\cos\theta_* =\pm \sqrt { 1- {H(r_\gamma)\over H(r_*)}}.
\end{equation}
Here the $-$ sign is to be used for $r_*>r_\gamma$, [implying $\theta_*\in(\pi/2,\pi)$], and the $+$ sign for $r_*<r_\gamma$, [implying $\theta_*\in(0,\pi/2)$]. 
For light emitted from the photon sphere itself, $r_*=r_\gamma$, we have $\theta_*=\pi/2$, hence both $\sin\theta=1$ and $\cos\theta_*=0$.
We can summarize the general situation as 
\begin{equation}
\cos\theta_* = -\sign(r_*-r_\gamma) \sqrt{1-  {H(r_\gamma)\over H(r_*)}}.
\end{equation}

\clearpage
--- A brief calculation, (see the introduction, and/or reference~\cite{phenomenology}), now shows that the \emph{solid angle} $\Delta\Omega_*$  subtended by the escape cone (for light emitted from $r_*$ and escaping to spatial infinity) is given (in \emph{steradians}) by \enlargethispage{20pt}
\begin{equation}\Delta\Omega_* = 2\pi[1-\cos\theta_* ] = 2\pi \left[ 1+ \sign(r_*-r_\gamma)  \sqrt{1-  {H(r_\gamma)\over H(r_*)}}\; \right].
\end{equation}
As $r_*\to r_\gamma$ we see $\Delta\Omega_*\to 2\pi$; so exactly half of that light will escape.
For an emission point located above the photon sphere more than half of the emitted light escapes, 
while for an emission point located below the photon sphere less than half of the emitted light escapes

\section{Optical metric}
The so-called ``\emph{optical metric}'' is defined (for static spacetimes presented in manifestly static form) by taking~\cite{optical}:
\begin{equation}
[g_{optical}]_{ab} = {g_{ab}\over |g_{tt}|}.
\end{equation}
Specifically, in terms of the definitions of the previous section, 
\begin{equation}(\d s^2)_{optical} = -\d t^2 + {B(r)\over A(r)}\; \d r^2 +  H(r) \;\d\Omega^2.
\end{equation}
The optical metric is (by construction) conformal to the physical metric --- it is in fact \emph{ultrastatic}~\cite{ultrastatic}, and has exactly the same null curves (and null geodesics) as the physical metric --- though the affine parameters may (and in general will) differ~\cite{affine}. 

Let us now define the ``optical area'' of the constant-$r$ spherical surfaces by
\begin{equation}\A(r) = 4\pi\;H(r).
\end{equation}
Then the global minimum of $H(r)$ is also a global minumum of the optical area $\A(r)$, and hence we see that  the photon sphere corresponds to a wormhole throat \emph{in the optical metric}, (not the physical metric).\footnote{And for multiple photons spheres, corresponding to multiple extrema of the optical area, one would be dealing with multiple extrema of the optical area --- typically both optical wormhole throats and optical wormhole anti-throats.}
Thence in terms of the optical area 
\begin{equation}\Delta\Omega_*
= 2\pi \left[ 1+\sign(r_*-r_\gamma) \sqrt{1-  {\A(r_\gamma)\over \A(r_*)}} \; \right].
\end{equation}
We shall soon interpret the optical area of the outermost photon sphere $\A(r_\gamma)$ in terms of the capture cross section for photons coming in from spatial infinity.

\section{Large-distance asymptotics}

Let us agree to use curvature coordinates for the physical metric
\begin{equation}\d s^2 = - A(r) \;\d t^2 + B(r) \;\d r^2+ r^2\{\d\theta^2+\sin^2\theta\,\d\phi^2\}.
\end{equation}
Then, since at large distances we can safely assume $r_*>r_\gamma$, we have
\begin{equation}\Delta\Omega_* = 2\pi \left[ 1+ \sqrt{1-  {r_\gamma^2 \;A(r_*)\over r_*^2 \;A(r_\gamma)}}\; \right].
\end{equation}
At large distances, assuming asymptotic flatness,  $A(r_*)\to 1+\O(1/r_*)$, so we have 
\begin{equation}\Delta\Omega_* = 4\pi - {\pi r_\gamma^2\over r_*^2\; A(r_\gamma)} + \O(1/r_*^3).
\end{equation} 
So almost everything escapes.

The capture cone is the complement of the escape cone:
\begin{equation}\Delta\Omega_{capture} = 4\pi - \Delta\Omega_* 
= {\pi \,r_\gamma^2\over r_*^2 \;A(r_\gamma)} + \O(1/r_*^3)
=  {\pi \,H(r_\gamma)\over r_*^2} + \O(1/r_*^3).
\end{equation}
This has a simple physical interpretation: The apparent cross section for photon capture, for photons coming in from spatial infinity, is
\begin{equation}
\sigma_{capture} = {\pi \,r_\gamma^2\over A(r_\gamma)}= \pi \,H(r_\gamma) = {\A(r_\gamma)\over 4}.
\end{equation}
(Note that this is given in terms of the optical area of the photon sphere, not the physical area of the horizon, so the superficial resemblance to the Bekenstein entropy is an illusion.)
Then
\begin{equation}\Delta \Omega_{capture} = {\sigma_{capture}\over r_*^2} + \O(1/r_*^3).
\end{equation}
Note that this relates large-distance asymptotic behaviour, $\sigma_{capture}$, to near-horizon physics in the form of the location of the circular photon orbits $r_\gamma$.

You can slightly refine this estimate for the escape cone solid angle by using one additional term in the PPN asymptotic flatness expansion. 
Write 
\begin{equation}A(r_*) = 1-2m_\infty/r_* + \O(1/r_*^2),
\end{equation} 
whence
\begin{equation}
\Delta\Omega_* = 4\pi - {\pi r_\gamma^2(1-2m_\infty/r_*)\over r_*^2 A(r_\gamma)} + \O(1/r_*^4).
\end{equation}
That is:
\begin{equation}
\Delta\Omega_* = 4\pi - {\sigma_{capture}(1-2m_\infty/r_*)\over r_*^2} + \O(1/r_*^4).
\end{equation}
In terms of the escape probability
\begin{equation}
P_* = 1 - {\sigma_{capture}\over 4\pi r_*^2} \; (1-2m_\infty/r_*) + \O(1/r_*^4).
\end{equation}
This is a quite general result, valid for photon emission at large distances.
All of the near-horizon physics has been hidden in the capture cross section $\sigma_{capture}$. 
The relative strength of the sub-leading $\O(1/r_*^3)$ large-distance term is controlled by the PPN expansion.

\section{Near-horizon physics: non extremal horizons}

Since sufficiently near the horizon we can safely assume $r_*<r_\gamma$, we have
\begin{equation}
\Delta\Omega_* = 2\pi \left[ 1- \sqrt{1-  {r_\gamma^2 \; A(r_*)\over r_*^2 \; A(r_\gamma)}} \;\right].
\end{equation}
At the horizon itself $A(r_H)=0$, so sufficiently near the horizon one has 
\begin{equation}A(r_*) = A'(r_H) \, [r_*-r_H] +\O([r_*-r_H]^2).
\end{equation}
Consequently
\begin{equation}\Delta\Omega_* =  {\pi r_\gamma^2  A'(r_H) \, [r_*-r_H] \over r_H^2 A(r_\gamma)} + \O([r_*-r_H]^2).
\end{equation}
So for near-horizon emission the escape cone indeed shrinks to zero, as expected. 

Now $ A'(r_H)$ has a direct physical  interpretation in terms of the surface gravity $\kappa_H$.  Let us, without loss of generality, set
\begin{equation}
A(r)=\exp[-2\Phi(r)] \; \{1-2m(r)/r\}; \qquad \hbox{and} \qquad B(r)={1\over1-2m(r)/r}.
\end{equation}
(Forcing $\Phi(r)\to0$ is an actual physical constraint on the metric~\cite{Jacobson}; there are situations where this is appropriate, [eg: Reissner--Nordstr\"om], but there are very many situations where this is not appropriate~\cite{not-Kiselev-1, not-Kiselev-2}.)

Then a standard calculation (see for instance reference~\cite{dirty})  yields
\begin{equation}
\kappa_H = \exp[-\Phi(r_H)] \; \left[\{1-2m(r)/r\}'\right]_H = \exp[+\Phi(r_H)] \; A'(r_H).
\end{equation}
Thence for non-extremal black holes we have the generic near-horizon result
\begin{equation}
\Delta\Omega_* =  {\pi r_\gamma^2  \;\exp[-\Phi(r_H)] \;\kappa_H \; [r_*-r_H] \over r_H^2 \;A(r_\gamma)} + \O([r_*-r_H]^2).
\end{equation}
That is
\begin{equation}
\Delta\Omega_* =  {\sigma_{capture}  \over r_H^2} \;\exp[-\Phi(r_H)] \;\kappa_H \; [r_*-r_H] + \O([r_*-r_H]^2).
\end{equation}
Furthermore in terms of the on-horizon energy density $\rho(r_H)$ we have~\cite{dirty}
\begin{equation}
\kappa_H = { \exp[-\Phi(r_H)]  \{ 1 - 8\pi \rho(r_H) r_H^2\} \over 2 \, r_H}.
\end{equation}
Thence
\begin{equation}
\Delta\Omega_* =  {\sigma_{capture}  \;\exp[-2\Phi(r_H)] \; \{ 1 - 8\pi \rho(r_H) r_H^2\}  \; [r_*-r_H] \over 2\, r_H^3} + \O([r_*-r_H]^2).
\end{equation} 
That is
\begin{equation}
\Delta\Omega_* =  {1\over2}\; {\sigma_{capture} \over  r_H^2} \;\exp[-2\Phi(r_H)] \; \{ 1 - 8\pi \rho(r_H) r_H^2\}  \; {[r_*-r_H] \over  r_H} + \O([r_*-r_H]^2).
\end{equation} 
So overall,  we can say quite a bit about the near-horizon behaviour of the escape cone, without having to be too explicit about the precise form of the spacetime geometry.

\section{Near-horizon physics: extremal horizons}

For extremal black holes one needs to go to at least one higher order in the Taylor series expansion of $A(r)$:
\begin{equation}
A(r_*) = {1\over2} A''(r_H) \, [r_*-r_H]^2 +\O([r_*-r_H]^3).
\end{equation}
Consequently
\begin{equation}
\Delta\Omega_* =  {\pi r_\gamma^2  \; A''(r_H) \; [r_*-r_H]^2 \over {2} \,r_H^2 \; A(r_\gamma)} + \O([r_*-r_H]^3).
\end{equation}
This can also be written as 
\begin{equation}
\Delta\Omega_* =  {\sigma_{capture} A''(r_H) \, [r_*-r_H]^2 \over 2 \,r_H^2 } + \O([r_*-r_H]^3).
\end{equation}
So for near-horizon emission from an extremal black hole the escape cone shrinks to zero even more rapidly than for a non-extremal black hole.

To get a physical interpretation for the quantity $ A''(r_H) $ for extremal black holes it is convenient to write the metric components in the form
\begin{equation}
A(r)=\exp[-2\Phi(r)] \; \{1-b(r)/r\}^2; \qquad \hbox{and} \qquad B(r)={1\over\{1-b(r)/r\}^2}.
\end{equation}
There is then a double root/double pole in the metric components, corresponding to an extremal horizon,  when $b(r_H)=r_H$. 

It is then easy to check that at the extremal horizon
\begin{equation}
G^t{}_t(r_H)= G^r{}_r(r_H) =- {1\over r_H^2}; \quad\hbox{and}\quad 
G^\theta{}_\theta(r_H) = G^\phi{}_\phi(r_H) = +{[1-b'(r_H)]^2\over r_H^2}.
\end{equation}
Furthermore a brief calculation yields
\begin{equation}
 A''(r_H)  = {2 e^{-2\Phi(r_H)} \;    [1-b'(r_H)]^2\over r_H^2} = 2 e^{-2\Phi(r_H)} \;  G^\theta{}_\theta.
\end{equation}
Thence in terms of the on-horizon transverse pressure $p_t(r_H)$ we have
\begin{equation}
\Delta\Omega_* =  {\sigma_{capture} \; e^{-2\Phi(r_H)} \;  \{8\pi p_t(r_H)\}  \, [r_*-r_H]^2 \over \,r_H^2 } + \O([r_*-r_H]^3).
\end{equation}
Finally, in terms of the on-horizon energy density $\rho(r_H)$ we have
\begin{equation}
\Delta\Omega_* =  {\sigma_{capture} \; e^{-2\Phi(r_H)} \;  \{8\pi \rho(r_H)\}  \, [1-b'(r_H)]^2\, [r_*-r_H]^2 \over  \,r_H^2 } + \O([r_*-r_H]^3).
\end{equation}
So again we see that quite a lot can be said, without having to be too explicit about the precise form of the spacetime geometry.

\section{Examples}

Let us now illustrate the general formalism developed above by considering  some standard (and not so standard) examples.
To keep the discussion relatively concrete and explicit we focus on black holes, wormholes, and ``black bounce'' spacetimes. Similar comments could be made concerning horizonless compact objects~\cite{RBH-2-ECO}.

\subsection{Black holes}

Let us  consider the Schwarzschild~\cite{Schwarzschild}, Kottler~\cite{Kottler}, and Reissner--Nordstr\"om~\cite{Reissner,Nordstrom} black holes.

\subsubsection{Schwarzschild}

For Schwarzschild spacetime:
\begin{equation}A(r)= 1-2m/r; \qquad D(r)=r^2; \qquad H(r) = {r^2\over 1-2m/r}; 
\end{equation}
\begin{equation}
r_\gamma=3m; \qquad H(r_\gamma) = 27 m^2;\qquad \sigma_{capture} = 27\pi m^2.
\end{equation}
Note there is only one photon sphere between the horizon and spatial infinity.
Then
\begin{equation}
\sin \theta_* = \sqrt{27} \;\; {m\over r_*} \;\sqrt{1-{2m\over r_*}};
\qquad
\cos \theta_* =  -\left(1-{3m\over r_*}\right)  \;\sqrt{1+{6m\over r_*}}.
\end{equation}
\begin{equation}
\Delta\Omega_* = 2\pi \left[ 1+ (1-3m/r_*)\sqrt{1+6m/r_*} \right].
\end{equation}
This reproduces the standard results, consistent with those reported in~\cite{Synge:1966, phenomenology}.\\
(The explicit $\pm$ in $\Delta\Omega_* $ cancels with the implicit $\pm$ coming from $\sqrt{x^2}=\pm x$.)

\subsubsection{Kottler (Schwarzschild--de Sitter)}

For the Schwarzschild--de~Sitter (Kottler) spacetime~\cite{Kottler}: 
\begin{equation}A(r)= 1-{2m\over r}- {\Lambda r^2\over 3}; \qquad D(r)=r^2; \qquad H(r) = {r^2\over 1-{2m\over r}-{\Lambda r^2\over3}}; 
\end{equation}
\begin{equation} 
r_\gamma=3m; \qquad H(r_\gamma) = {27 m^2\over 1 - {9\Lambda m^2}};
 \qquad \sigma_{capture} = {27 \pi m^2\over 1 - {9\Lambda m^2}};
\end{equation}
Note for de Sitter space there is only one photon sphere between the outermost black hole horizon 
and the cosmological horizon, 
or in the case of anti--de Sitter space, only one photon sphere between the outermost black hole horizon 
horizon and spatial infinity.
Then 
\begin{equation}
\Delta\Omega_* = 2\pi \left[ 1+ \sign(r_*-3m)  \sqrt{1 - {27m^2\over r_*^2} \;
{1-{2m\over r_*}-{\Lambda r_*^2\over3} \over 1 -9 \Lambda m^2} }\right].
\end{equation}
This can be rearragnged to finally yield
\begin{equation}
\Delta\Omega_* = 2\pi \left[ 1+ {(1-3m/r_*)\sqrt{ 1+6m/r_*} \over \sqrt{1 -9 \Lambda m^2}} \right].
\end{equation}
So going from Schwarzschild to Kottler merely requires strategically introducing the constant factor $(1-9\Lambda m^2)$ in a few key places. 
Note that as $r_*$ approaches the black hole horizon we have $\Delta\Omega_*\to0$, whereas as $r_*\to 3m$ we have $\Delta\Omega_*\to2\pi$. 

For asymptotically de Sitter space ($\Lambda>0$), as  $r_*$ approaches the cosmological horizon, (where in particular $1-{2m\over r_*}-{\Lambda r_*^2\over3} \to 0$), we have $\Delta\Omega_*\to4\pi$. In contrast for asymptotically anti-de Sitter space ($\Lambda<0$) there is no cosmological horizon, and as  $r_*\to\infty$  we have the perhaps somewhat unexpected result
\begin{equation}
\Delta\Omega_* \to 2\pi \left[ 1+ {1\over \sqrt{1 +9\, |\Lambda|\, m^2}} \right] < 4\pi.
\end{equation}

\subsubsection{Reissner--Nordstr\"om}
\paragraph{General case:}

For the Reissner--Nordstr\"om spacetime the general situation is
\begin{equation}
A(r)= 1-{2m\over r}+ {Q^2\over r^2}; \qquad D(r)=r^2; \qquad H(r) = {r^2\over 1-{2m\over r}+{Q^2\over r^2}}; 
\end{equation}
\begin{equation}
r_\gamma={3m\over2} \left[1 +\sqrt{1 -{ 8Q^2\over9 m^2}}\;\right] 
= {3m\over2} +\sqrt{{9m^2\over4}-Q^2} \leq 3m.
\end{equation}
The outer horizon is located at
\begin{equation}
r_H=m +\sqrt{m^2 -Q^2} < r_\gamma.
\end{equation}
Note $r_\gamma>r_H$, and so there is only one photon sphere between the horizon and spatial infinity.

Then the explicit formula for $H(r_\gamma)$ is a bit of a mess, but we can certainly say
\begin{eqnarray}
\Delta\Omega_* 
&=& 
2\pi \left[ 1 +\sign(r_*-r_\gamma)  \sqrt{ 1- \left({r_\gamma\over r_*}\right)^2 {1-{2m\over r_*}+{Q^2\over r_*^2}\over 
1-{2m\over r_\gamma}+{Q^2\over r_\gamma^2}}}\;
 \right]
 \nonumber\\
 &=&
 2\pi \left[ 1 +\sign(r_*-r_\gamma) \sqrt{ 1-  \left({r_\gamma\over r_*}\right)^4 {r_*^2-{2m r_*}+{Q^2}\over 
r_\gamma^2-{2m r_\gamma}+{Q^2}}}\;
 \right].
\end{eqnarray}
We can also write this as 
\begin{equation}
\Delta\Omega_*  = 
2\pi \left[ 1 +\sign(r_*-r_\gamma)  \sqrt{ 1- \left(\sigma_{capture}\over \pi r_*^2\right) 
\left(1-{2m\over r_*}+{Q^2\over r_*^2}\right) }\;\right].
\end{equation}
By inspection, as $r_*$ approaches the (outer) horizon $\Delta\Omega_*\to 0$, and 
as $r_*$ approaches the photon sphere $\Delta\Omega_*\to 2\pi$. At large distances we recover
\begin{equation}
\Delta\Omega_*  =  4\pi - \left(\sigma_{capture}\over  r_*^2\right) 
\left(1-{2m\over r_*}\right)  +\O(1/r_*^4).
\end{equation}

\paragraph{Near-horizon behaviour:}

Near the outer horizon of a Reissner--Nordstr\"om black hole we easily see
\begin{equation}
\left(1-{2m\over r_*}+{Q^2\over r_*^2}\right)  
= \kappa_H  [r_*-r_H] + \O([r_*-r_H]^2).
\end{equation}
Thence (as expected)
\begin{equation}
\Delta\Omega_* = {\sigma_{capture}\over r_H^2}\; \kappa_H \;  [r_*-r_H] + \O([r_*-r_H]^2).
\end{equation}

\paragraph{Small charge:}

\def\O{{\mathcal{O}}}
For small charge, ($|Q|\ll m$), we have
\begin{equation}
r_\gamma=3m \left[ 1 + {2\over9}{Q^2\over m^2} -{4\over81} {Q^4\over m^4} + \O\left(Q^6\over m^6\right) \right];
\end{equation}
and
\begin{equation}H(r_\gamma) = 27 m^2 \left[ 1 - {1\over3}{Q^2\over m^2} -{1\over27} {Q^4\over m^4} 
+ \O\left(Q^6\over m^6\right) \right].
\end{equation}

Thence
\begin{eqnarray}
\Delta\Omega_* 
&=& 2\pi \left[ 1- (1-3m/r_*)\sqrt{1+6m/r_*} \right]
+ {9\pi (m^2/r_*^2) (1+m/r_*)\over \sqrt{1+6m/r_*}}\; {Q^2\over m^2} 
\nonumber\\
&&
+ 
{\pi (m^2/r_*^2) (1+{7m\over r_*} -{9\over4}{m^2\over r_*^2} +{27\over 4} {m^3\over r_*^3}) \over (1+6m/r_*)^{3/2} }\; {Q^4\over m^4} 
+ \O\left(Q^6\over m^6\right).
\end{eqnarray}
Note that these results are the Schwarzschild results with $\O(Q^2/m^2)$ modifications.

\paragraph{Extremal limit $m=|Q|$:}
For the extremal Reissner--Nordstr\"om spacetime
\begin{equation}A(r)= \left(1-{m\over r}\right)^2; \qquad D(r)=r^2; \qquad H(r) = {r^2\over (1-{m\over r})^2}; \qquad r_\gamma=2m.
\end{equation}
Thence
\begin{equation}
H(r_\gamma)= 16 m^2; \qquad \sigma_{capture}=16 \pi m^2;
\end{equation}
and
\begin{equation}
\sin\theta_* = 4\; {m\over r_*} \left(1-{m\over r_*}\right); \qquad 
\cos\theta_* = -\left(1-{2m\over r_*}\right)\sqrt{1+{4m\over r_*}-{4m^2\over r_*^2}}.
\end{equation}
Finally
\begin{equation}
\Delta\Omega_* 
 =
 2\pi \left[ 1+ \left(1-{2m\over r_*}\right)\sqrt{1+{4m\over r_*}-{4m^2\over r_*^2}} \; \right].
\end{equation}

At large distances, as expected, we find
\begin{equation}
\Delta\Omega_* 
 = 4\pi - {16 \pi m^2\over r_*^2}\left(1-{2m\over r_*}\right) +  \O(1/r_*^4)
\end{equation}

Near the horizon (now at $r_H=m$) we have, as expected,
\begin{equation}
\Delta\Omega_* 
= 16\pi \left(r_*-m\over m\right)^2 + \O([r-m]^3).
\end{equation}

\subsection{Traversable wormholes}
Traversable wormholes have attracted considerable theoretical attention over the last 35 years~\cite{Morris-Thorne,MTY,examples,surgery,book,small,exponential}, with even a few observational tests~\cite{natural, Zhou:2016, Takahashi:2013, Bambi:2021}. Let us consider three simple examples.

\subsubsection{Canonical Morris--Thorne wormhole}

The canonical Morris--Thorne wormhole is represented by the line element~\cite{Morris-Thorne}:
\begin{equation}
\d s^2 = -\d t^2 + \d\ell^2 + (a^2+\ell^2)\; \d\Omega^2; \qquad \ell\in(-\infty,+\infty).
\end{equation}
For this metric
\begin{equation}
H(\ell)=a^2+\ell^2.
\end{equation}
The location $\ell_\gamma$ of the photon sphere is then given by
\begin{equation}
\left.\frac{dH(\ell)}{d\ell}\right|_{\ell=\ell_{\gamma}}=0,
\end{equation}
which gives
\begin{equation}
\ell_{\gamma}=0.
\end{equation}
That is, the unique photon sphere for this spacetime is simply the wormhole throat itself. Note
\begin{equation}
H(\ell_\gamma)=a^2; \qquad \sigma_{capture} = \pi a^2.
\end{equation}

Then
\begin{equation}
\sin\theta_*  = \sqrt{\frac{H(\ell_{\gamma})}{H(\ell_*)}} = \frac{a}{\sqrt{a^2+\ell_*^2}}; \qquad
\cos\theta_*  = -\frac{|\ell_*|}{\sqrt{a^2+\ell_*^2}};
\end{equation}
(The signs are chosen so that far from the wormhole everything escapes into the same universe the emission point is in.)

So we have the rather simple result
\begin{equation}
\Delta\Omega_*  =2\pi\left(1+\sqrt{1-\frac{H(\ell_{\gamma})}{H(\ell_*)}}\right)\\
 = 2\pi\left(1+\frac{|\ell_*|}{\sqrt{a^2+\ell_*^2}}\right).
\end{equation}
Far from the wormhole throat (in either universe) $\Delta\Omega_* \to 4\pi$. As one approaches the wormhole throat (from either direction) $\Delta\Omega_* \to 2\pi$, indicating that half of the photons escape into each of the two asymptotic regions. 
For large $r_*$ we note that sine $m_\infty=0$ one has
\begin{equation}
\Delta\Omega_* = 4\pi - {\pi a^2\over \ell_*^2} + \O(1/\ell_*^4).
\end{equation}

\subsubsection{Exponential wormhole}

The exponential wormhole~\cite{exponential} is a particularly interesting toy model because the two sides of the wormhole throat are very grossly dissimilar. The line element is (in isotropic coordinates)
\begin{equation}
\d s^2 = -e^{-2m/r} \d t^2 + e^{+2m/r} \{  \d r^2+ r^2 \d \Omega^2 \}; \qquad r\in(0,\infty).
\end{equation}
The global minimum of $r^2 e^{+2m/r}$ defines a wormhole throat, located at $r=m$. 
Note the ``origin'' $r=0$ is an infinite proper distance from the throat: $\int_0^m e^{+m/r} dr = \infty$. \\
Furthermore, the volume of the interior region $0\leq r\leq m$ is also infinite. Indeed:  $4\pi \int_0^m r^2 e^{+3m/r} dr = \infty$.
However, we emphasize that the region ``below'' the throat is both qualitatively and quantitatively rather different from the region ``above'' the throat~\cite{exponential}; the wormhole is very definitely asymmetrical. 
We have
\begin{equation}
A(r) = e^{-2m/r}; \qquad B(r) = e^{+2m/r}; \qquad D(r) = r^2 e^{+2m/r}; \qquad H(r) = r^2 e^{+4m/r}.
\end{equation}
The relevant optical metric is 
\begin{equation}
(\d s^2)_{optical} = -\d t^2 + e^{+4m/r} \{  \d r^2+ r^2 \d \Omega^2 \}; \qquad r\in(0,\infty).
\end{equation}
Then the circular photon orbits are defined by the extrema of $ r^2 e^{+4m/r}$, equivalently the extrema of $r e^{+2m/r}$,  and are unique, being located at $r_\gamma=2m$.
Thence
\begin{equation}
H(r_\gamma) = 4 m^2 e^2; \qquad \hbox{and} \qquad \sigma_{capture} = 4\pi m^2 e^2.
\end{equation}
So for the escape cone for emission from $r_*$ into the region $r>2m$ we have
\begin{equation}
\sin\theta_* = {2m\over r_*}  \; e^{1-2m/r_*}; \qquad \cos\theta_* = -\sign(r_*-2m) \sqrt{1-  {4m^2\over r_*^2} \; e^{2(1-2m/r_*)}}.
\end{equation}
Finally
\begin{equation}
\Delta\Omega_* = 2\pi \left[ 1 + \sign(r_*-2m) \sqrt{1-  {4m^2\over r_*^2} \; e^{2(1-2m/r_*)}}\;\right].
\end{equation}
For emission from $r_*\gg m$ we have
\begin{equation}
\Delta\Omega_* = 4\pi- {4\pi m^2 e^2\over r_*^2} \left[ 1 - {4m\over r_*}\;\right] +\O(1/r_*^4).
\end{equation}
Because the throat at $r=m$ is displaced from the photon sphere at $r_\gamma=2m$, photon emission from an isotropic source placed at the physical wormhole throat into the ``outside'' will be asymmetrical. Specifically
\begin{equation}
[\Delta\Omega_*]_{throat} = 2\pi \left[ 1 - \sqrt{1-  {4\over e^2}}\;\right] \approx 2\pi \times (0.3227564196...).
\end{equation}

For emission from $r \ll m$, deep in the ``interior'' of the wormhole,  we have the exponentially narrow escape cone
\begin{equation}
\Delta\Omega_* = {4\pi m^2 e^2\over r_*^2}  \; e^{-4m/r_*} +\O(e^{-8m/r_*}/r_*^4).
\end{equation}

\subsubsection{Generic Morris--Thorne wormhole: Single photon sphere}

The generic Morris--Thorne wormhole in proper distance coordinates is represented by the line element~\cite{MTY}:
\begin{equation}\label{E:proper}
\d s^2 = - e^{-2\Phi(\ell)} \d t^2 + \d\ell^2 + r(\ell)^2 \d\Omega^2; \qquad \ell\in(-\infty,+\infty).
\end{equation}
One sometimes sees this generic wormhole represented in curvature coordinates, which are a little trickier because one has to patch two coordinate charts together~\cite{Morris-Thorne}:
\begin{equation}
\d s^2 = - e^{-2\Phi(r)} \;\d t^2 +{\d r^2\over 1-b(r)/r} + r^2 \;\d\Omega^2; \qquad r\in(a,+\infty); \quad b(a) =a.
\end{equation}
So let us agree to work in proper distance coordinates. 
For the metric (\ref{E:proper})  we find
\begin{equation}
H(\ell)= e^{2\Phi(\ell)} \; r(\ell)^2.
\end{equation}
Since by construction the wormhole is assumed to have no horizons, 
and in the two asymptotic regions $H(\ell)\approx \ell^2 +\O(1)$, 
there will be a global minimum of $H(\ell)$. But there may be \emph{many} extrema. 
For now let us assume that the photon sphere is unique. 
Then
\begin{equation}
\left.\frac{dH(\ell)}{d\ell}\right|_{\ell=\ell_{\gamma}}=0,
\end{equation}
has the implicit solution
\begin{equation}
\Phi'(\ell_\gamma)+ {r'(\ell_\gamma)\over r(\ell_\gamma)} = 0.
\end{equation}
In contrast the wormhole throat is located at the  minimum of $r(\ell)$. 
That is, the photon sphere for this generic wormhole spacetime need not be in the same location as the  the wormhole throat.\footnote{We have already seen this happen for the exponential wormhole.} 

For the escape cone opening angle, for emission from $\ell_*$ into the asymptotic region $\ell\gg \ell_\gamma$,  we have
\begin{equation}
\sin\theta_*  = \sqrt{\frac{H(\ell_{\gamma})}{H(\ell_*)}}
 = \frac{r(\ell_{\gamma})}{r(\ell_*)}\;\exp[\Phi(\ell_*)-\Phi(\ell_{\gamma})].
\end{equation}
Thence
\begin{eqnarray}
\cos\theta_* &=& -\sign(\ell_*-\ell_\gamma) \sqrt{1-\frac{H(\ell_{\gamma})}{H(\ell_*)}}\;
\nonumber\\
&=& -\sign(\ell_*-\ell_\gamma)\sqrt{1-\frac{r(\ell_{\gamma})^2}{r(\ell_*)^2}\;\exp[2\Phi(\ell_*)-2\Phi(\ell_{\gamma})]}.
\end{eqnarray}
So for the escape cone solid angle we have
\begin{eqnarray}
\Delta\Omega_* &=& 2\pi\left[1+\sign(\ell_*-\ell_\gamma) \sqrt{1-\frac{H(\ell_{\gamma})}{H(\ell_*)}}\;\right]
\nonumber\\
&=& 2\pi\left[1+\sign(\ell_*-\ell_\gamma)\sqrt{1-\frac{r(\ell_{\gamma})^2}{r(\ell_*)^2}\;\exp[2\Phi(\ell_*)-2\Phi(\ell_{\gamma})]}\;\right].
\end{eqnarray}
For emission at large distances above the photon orbit $\Delta\Omega_* \to 4\pi$, while on the photon orbit  itself $\Delta\Omega_*\to 2\pi$. To say anything more specific one would need to choose a particular model for both $r(\ell)$ and $\Phi(\ell)$.

\subsubsection{Multiple photon spheres}

If one is dealing with multiple photon spheres, then extra caution is called for. 
In the words of reference~\cite{Perlick:2021}: ``If there are several photon spheres, [and one is considering silhouettes], one has to specify where precisely one assumes [the background] light sources to be situated.'' 

Likewise, if there are several photon spheres, [and one is considering escape cones to spatial infinity], one has to specify where precisely one assumes the emission point to be situated relative to these multiple photon spheres. This point is important when considering  truly generic multi-photon-sphere versions of the Morris--Thorne wormholes, 
 and will also show up below when we consider a specific sub-case of the ``black bounce'' spacetimes. 

If one is dealing with multiple photon spheres, then one can construct multiple ``naive escape cones'', one for each photon sphere --- these ``naive escape cones'' would be defined by those null curves that asymptote to the relevant photon sphere. 
The true escape cone would then be defined by the intersection of these ``naive escape cones''. 
That is, the true escape cone would then be defined by the  smallest of these ``naive escape cones'', corresponding to the smallest value of $H(r_\gamma)$ . 

\begin{itemize}
\item 
If the emission point $r_*$ lies below at most one of the photon spheres then this is simple --- the physics of escape is controlled by looking for the outermost photon sphere (either above or below $r_*$), which in view of the assumed asymptotic flatness of the spacetime, must be a local minimum of $H(r)$, and so correspond to an unstable photon sphere. So the usual calculation applies.
\item
If the emission point $r_*$ lies below two or more of the photon spheres, then the discussion becomes more complex --- not all of the photon spheres can be unstable. Indeed, at least one of the photon spheres must be at a local maximum of $H(r)$ and so correspond to a stable photon orbit. The dichotomy between \emph{escape} and \emph{capture} then becomes a trichtomy, between  \emph{escape} (to spatial infinity), \emph{capture} (by the black hole or the universe on the other side of the wormhole), and localized \emph{trapping} by the stable photon sphere.  
\end{itemize}

In general this is all that can be said without specifying a more explicit model --- fortunately an explicit tractable example of this behaviour will show up below when we consider a specific sub-case of the ``black bounce'' spacetimes.

\subsection{Black bounce spacetimes}\enlargethispage{20pt}
The black bounce spacetimes have a tuneable parameter $a$ that allows one (among other things) to interpolate between black holes and wormholes. The simplest canonical example is~\cite{Simpson:2018}:
\begin{equation}
\d s^{2}=-\left(1-\frac{2m}{\sqrt{r^{2}+a^{2}}}\right)\d t^{2}+\frac{dr^{2}}{1-\frac{2m}{\sqrt{r^{2}+a^{2}}}}
+\left(r^{2}+a^{2}\right)\left(\d\theta^{2}+\sin^{2}\theta \;\d\phi^{2}\right),
\end{equation}
with $r\in(-\infty,+\infty)$ and $a\geq 0$. 
(Note that very many other variations on this theme are possible~\cite{Lobo:2020a, Lobo:2020b, Franzin:2021}.)
If $a<2m$ then this geometry describes a regular black hole, with horizons at $r_H^\pm=\pm\sqrt{(2m)^2-a^2}$. 
If $a>2m$ this geometry describes a traversable wormhole with throat at $r=0$. 

In terms of the notation developed above
\begin{equation}
A(r)= 1-\frac{2m}{\sqrt{r^{2}+a^{2}}}; \qquad B(r) = {1\over 1-\frac{2m}{\sqrt{r^{2}+a^{2}}}}; \qquad D(r)= r^2+a^2;
\end{equation}
and so
\begin{equation}
H(r)= {r^2+a^2\over 1-\frac{2m}{\sqrt{r^{2}+a^{2}}}}.
\end{equation}
The relevant optical metric is
\begin{equation}
(\d s^{2})_{optical}=-\d t^{2}+\frac{\d r^{2}}{\left(1-\frac{2m}{\sqrt{r^{2}+a^{2}}}\right)^2}
+\left(r^{2}+a^{2}\over1-\frac{2m}{\sqrt{r^{2}+a^{2}}} \right)\left(\d\theta^{2}+\sin^{2}\theta \;\d\phi^{2}\right),
\end{equation}

\begin{itemize}
\item 
For black holes ($a<2m$) there are 2 photon spheres at $r_\gamma=\pm \sqrt{(3m)^2-a^2}$.   For both of these photon spheres 
\begin{equation}
H(r_\gamma^\pm) = 27 m^2; \qquad (a< 2m),
\end{equation}
which is just the result for Schwarzschild. 
Furthermore
\begin{equation}
H''(r_\gamma^\pm) = 18 - 2(a^2/m^2) >10>0;  \qquad (a< 2m),
\end{equation}
so as expected these are both local minima corresponding to unstable photon spheres. 
While formally $H'(r)=0$ has a third root at $r=0$ this now occurs in a region where the $r$ coordinate is timelike, so this root does not correspond to a photon sphere. 
\item
For wormholes with the parameter $a$ in the range $a\geq3m$ there is only one photon sphere, located at the wormhole throat $r_\gamma^0=0$. We then have
\begin{equation}
H(r_\gamma^0) = {a^3\over a-2m} = {a^2\over 1-2m/a}; \qquad (a\geq 3m).
\end{equation}
Furthermore
\begin{equation}
H''(r_\gamma^0) = {2a(a-3m)\over(a-2m)^2 } >0; \qquad (a\geq 3m).
\end{equation}
So the photon sphere at $r_\gamma^0 =0$ is unstable.
It is this specific case that is qualitatively most similar to the canonical Morris--Thorne wormhole (which would correspond to the limit $m\to0$.)
\item 
For wormholes with the parameter $a$ in the range $a\in(2m,3m)$ there are \emph{three} photon spheres, two at $r_\gamma=\pm \sqrt{(3m)^2-a^2}$ \emph{and} one at the wormhole throat $r_\gamma^0=0$. 

In this parameter regime we still have 
\begin{equation}
H(r_\gamma^\pm) = 27 m^2; \qquad  H''(r_\gamma^\pm) = 18 - 2(a^2/m^2) >0;\qquad (2m<a <3m),
\end{equation}
but now 
\begin{equation}
H(r_\gamma^0) = {a^2\over 1-2m/a} \geq 27 m^3 ; \quad H''(r_\gamma^0) = {2a(a-3m)\over(a-2m)^2}<0;\quad (2m<a <3m).
\end{equation}
So in this parameter regime the two photon spheres at $r_\gamma^\pm$ are unstable, 
while the third photon sphere at the wormhole throat is stable.
\end{itemize}
To analyze the escape cones it is best to consider these three cases separately.

\subsubsection{Black bounce regular black hole}

In the black hole situation, $2m>a$, one might as well restrict attention to one asymptotic region $r\geq r_H= \sqrt{(2m)^2-a^2}$ and its unique corresponding photon sphere $r_\gamma=\sqrt{(3m)^2-a^2}$. Then for isotropic photon emission from $r_*\geq r_H$ the escape cone is 
\begin{equation}
\sin\theta_* = \sqrt{27 m^2 \left(1-\frac{2m}{\sqrt{r_*^{2}+a^{2}}}\right)
\over r_*^2+a^2}.
\end{equation}
After a little bit of work
\begin{equation}
\cos\theta_* = - \left(1-{3m\over\sqrt{r_*^2+a^2} }\right) \sqrt{1+{6m\over\sqrt{r_*^2+a^2}}}.
\end{equation}
(It is easy to see that this has the appropriate Schwarzschild limit as $a\to0$.) 

Then for the escape cone solid angle
\begin{equation}
\Delta\Omega_* = 2\pi\left[ 1 + \left(1-{3m\over\sqrt{r_*^2+a^2} }\right) \sqrt{1+{6m\over\sqrt{r_*^2+a^2}}}\;\right].
\end{equation}

At large distances $r_*\gg m$ we find
\begin{equation}
\Delta\Omega_* = 4\pi - {27\pi m^2\over r_*^2} (1-2m/r_*) + \O(1/r^4).
\end{equation}
This is (asymptotically) the same as the Schwarzschild result, as the $a$ dependence first shows up at the next higher order $\O(a^2 m^2/r_*^4)$.

The near-horizon limit is most transparently given in terms of $r_H$ as 
\begin{equation}
\Delta\Omega_* = {27\pi\, r_H\over 16 \,m^2} (r_*-r_H) +\O([r_*-r_H]^2). 
\end{equation}
All of this agrees with our general analysis.

\subsubsection{Black bounce wormhole: single photon sphere}

For $a\geq 3m$ there is only a single photon sphere, located at the wormhole throat $r=0$. 
Thence applying out general framework
\begin{equation}
\sin\theta_* = \sqrt{a^2 \left(1-\frac{2m}{\sqrt{r_*^{2}+a^{2}}}\right)
\over (1-2m/a)(r_*^2+a^2)}
=
{a \over \sqrt{r_*^2+a^2}} \; \sqrt{1-\frac{2m}{\sqrt{r_*^{2}+a^{2}}}
\over 1-2m/a}.
\end{equation}
Thence
\begin{equation}
\cos\theta_* = -\;\sign(r_*)\;\sqrt{ 1 - {a^2\over r_*^2+a^2} \; {1-\frac{2m}{\sqrt{r_*^{2}+a^{2}}}\over1-2m/a}}.
\end{equation}
Then for the escape cone solid angle (for emission from $r_*$ into the $r>0$ asymptotic region) we see
\begin{equation}
\Delta\Omega_* = 2\pi\left[1+ \sign(r_*)\;\sqrt{ 1 - {a^2\over r_*^2+a^2} \; {1-\frac{2m}{\sqrt{r_*^{2}+a^{2}}}\over1-2m/a}}\;\right].
\end{equation}
For the complementary capture cone solid angle (for emission from $r_*$ into the $r<0$ asymptotic region) we see
\begin{equation}
(\Delta\Omega_*)_{capture} = 2\pi\left[1- \sign(r_*)\;\sqrt{ 1 - {a^2\over r_*^2+a^2} \; {1-\frac{2m}{\sqrt{r_*^{2}+a^{2}}}\over1-2m/a}}\;\right].
\end{equation}

For an isotropic source at asymptotically large $r_*\gg 2m$ we have
\begin{equation}
\Delta\Omega_* = 4\pi - {\pi a^2\over r_*^2 (1- 2m/a)} \left(1-{2m\over r_*} \right)
+\O(1/r_*^4)
\end{equation}

For an isotropic source at the wormhole throat $r_*=0$ we have $\Delta\Omega_*=2\pi$. (This half-half division of the emitted photons is exactly what we expect based on the $+r\leftrightarrow -r$ symmetry in the line element.)

\clearpage
For an isotropic source at asymptotically large negative $r_*=-|r_*| \ll -2m$ , (that is, far away on the other side of the wormhole throat), and for the case of emission into the positive $r$ asymptotic region, we have
\begin{equation}
\Delta\Omega_* = +{\pi a^2\over r_*^2 (1- 2m/a)} \left(1-{2m\over |r_*|} \right).
+\O(1/r_*^4)
\end{equation}
This is again what we expect based on the explicit $+r\leftrightarrow -r$ symmetry in the line element; emitted 
photons will evenutually arrive at either one or the other asymptotic regions.

\subsubsection{Black bounce wormhole: triple photon sphere}

When $a\in(2m,3m)$ there are now 3 photon spheres,  located at $r_\gamma^\pm=\pm\sqrt{(3m)^2-a^2}$ and $r_\gamma^0=0$ respectively, 
and a correspondingly more complex set of questions one can in principle ask about escape cones.

\paragraph{Sources on our side of the wormhole throat:}
Let us first focus on sources and  observers in ``our'' asymptotic region ($r_*>0$ and $r\gg0$).
 That is, we allow $r_*\in(0,+\infty)$, focus on the effect of the single photon sphere $r_\gamma^+=\sqrt{(3m)^2-a^2}$, 
and consider emission into the region $r\gg0$. 
Then for isotropic photon emission from $r_*>0$ the escape cone is formally the same as for the black hole case, though now one has $a\in(2m,3m)$:
\begin{equation}
\sin\theta_* = \sqrt{27 m^2 \left(1-\frac{2m}{\sqrt{r_*^{2}+a^{2}}}\right)
\over r_*^2+a^2}.
\end{equation}
After a little bit of work
\begin{equation}
\cos\theta_* = - \sign\left(r_* - \sqrt{(3m)^2-a^2}\right) 
\left|1-{3m\over\sqrt{r_*^2+a^2} }\right|\; \sqrt{1+{6m\over\sqrt{r_*^2+a^2}}}.
\end{equation}
Thence for $r_*>0$ but otherwise unconstrained
\begin{equation}
\cos\theta_* = - 
\left(1-{3m\over\sqrt{r_*^2+a^2} }\right)\; \sqrt{1+{6m\over\sqrt{r_*^2+a^2}}}.
\end{equation}
(It is easy to see that this has the appropriate Schwarzschild limit as $a\to0$.) 

Then for $r_*>0$ the escape cone solid angle is 
\begin{equation}
\Delta\Omega_* = 2\pi\left[ 1 + 
 \left(1-{3m\over\sqrt{r_*^2+a^2} }\right)\; \sqrt{1+{6m\over\sqrt{r_*^2+a^2}}}\;\right].
\end{equation}

At large distances $r_*\gg m$ we find
\begin{equation}
\Delta\Omega_* = 4\pi - {27\pi m^2\over r_*^2} (1-2m/r_*) + \O(1/r_*^4).
\end{equation}
This is again (asymptotically) the Schwarzschild result, as the $a$ dependence first shows up at the next higher order $\O(a^2 m^2/r_*^4)$.
There is no near-horizon limit since there are no horizons.
On the other hand as $r_*\to r_\gamma^+$ we do have $\Delta\Omega_*\to 2\pi$, as expected. 

For emission from a point infinitesimally above the wormhole throat, but still in ``our'' universe, (so $r_*=0^+$), we have
\begin{equation}
(\sin\theta_*)_{throat} = \sqrt{27} \; {m\over a} \;  \sqrt{1-\frac{2m}{a}}.
\end{equation}
\begin{equation}
(\cos\theta_*)_{throat} = - 
\left(1-{3m\over a} \right)\; \sqrt{1+{6m\over a}}.
\end{equation}
\begin{equation}
(\Delta\Omega_*)_{throat} = 2\pi\left[ 1 +
 \left(1-{3m\over a }\right) \sqrt{1+{6m\over a}}\;\right].
\end{equation}
Since $a\in(2m,3m)$ and in particular $a<3m$ this is better written as 
\begin{equation}
(\Delta\Omega_*)_{throat} = 2\pi\left[ 1 -
 \left({3m\over a }-1\right) \sqrt{1+{6m\over a}}\;\right] < 2\pi.
\end{equation}
As $a\to3m$, (so that the throat merges with the outer photon sphere), we see $(\Delta\Omega_*)_{throat}\to 2\pi$. 
As $a\to2m$, (so that the geometry develops a horizon), we see $(\Delta\Omega_*)_{throat}\to 0$. 
\enlargethispage{20pt}

For $4\pi$ isotropic emission  from the wormhole throat itself we expect half the photons to initially move to $r>0$ and half to $r<0$. But we have just seen that for escape to positive infinity $(\Delta\Omega_*)_{throat} <2\pi$, and by symmetry there will be an equal escape cone to negative infinity. Thence there must be a certain fraction of the photons that neither make it out to positive infinity nor negative infinity --- ultimately being trapped by the presence of the stable photon sphere. Indeed
\begin{equation}
(\Delta\Omega_*)_{throat}^{trapping} = 4\pi - 2 (\Delta\Omega_*)_{throat}  =
  4\pi \left({3m\over a }-1\right) \sqrt{1+{6m\over a}}.
\end{equation}
This quantity is positive and less than $4\pi$ over the entire range $a\in(2m,3m)$.\\
 It tends to zero as $a\to3m$ and tends to unity as $a\to2m$. 
We shall soon see this trapping behaviour  generalize for emission from anywhere in the region between the two unstable photon spheres.

\paragraph{Sources between the unstable photon spheres:}
We have already dealt with the case $0<r_*<r_\gamma^+$. Let us now consider emission from the region $r_\gamma^-< r_* < 0$ into the region $r\gg r_\gamma^+$. Any such photon would have to cross two photon spheres, the stable photon sphere at the wormhole throat and the unstable photon sphere at $r_\gamma^+$. But in this situation
\begin{equation}
\min\{ H(r_\gamma)\} = \min\{H(r_\gamma^0),H(r_\gamma^-)\} 
= \min\left\{ {a^2\over 1-2m/a}, 27 m^2\right \} = 27 m^2.
\end{equation}
Thence it is the photon sphere at $r_\gamma^+$ that is determinative in controlling escape to $r\gg r_\gamma^+$, and the usual calculation for $r_*>0$ still applies:
\begin{equation}
\Delta\Omega_* = 2\pi\left[ 1 + 
 \left(1-{3m\over\sqrt{r_*^2+a^2} }\right)\; \sqrt{1+{6m\over\sqrt{r_*^2+a^2}}}\;\right].
\end{equation}
In counterpoint for emission from the region $r_\gamma^-< r_* < 0$ into the region $r\ll r_\gamma^-$, symmetry implies that this must be equivalent to emission from $|r_*|$ into the region $r\gg r_\gamma^+$ so we again have
\begin{equation}
\Delta\Omega_* = 2\pi\left[ 1 + 
 \left(1-{3m\over\sqrt{r_*^2+a^2} }\right)\; \sqrt{1+{6m\over\sqrt{r_*^2+a^2}}}\;\right].
\end{equation}
But since the escape cones into the two asymptotic regions does not add up to $4\pi$ steradians, this implies that some fraction of photons must remain trapped between the two unstable photon spheres. Specifically:
\begin{equation}
(\Delta \Omega_*)^{trapping}= 4\pi - 2 \Delta\Omega_*,
\end{equation}
and so
\begin{equation}
(\Delta \Omega_*)^{trapping}= 4\pi \; \left({3m\over\sqrt{r_*^2+a^2} }-1\right)\; \sqrt{1+{6m\over\sqrt{r_*^2+a^2}}}.
\end{equation}
Note that as $|r_*|\to|r_\gamma^\pm|$ we have $(\Delta \Omega_*)^{trapping}\to 0$, while as $r_*\to 0$ we recover our previous result
$(\Delta \Omega_*)^{trapping}\to (\Delta \Omega_*)^{trapping}_{throat}$. 

Another way of analyzing the situation is to note that if $|r_*|< |r_\gamma^\pm|$, so that you are between the unstable photon spheres, then the trapped region is defined by considering the photon trajectories satisfying
\begin{equation}
\sin\theta > \sin\theta_* = \sqrt{27 m^2 \left(1-\frac{2m}{\sqrt{r_*^{2}+a^{2}}}\right)
\over r_*^2+a^2}.
\end{equation}
and
\begin{equation}
-\cos\theta_* < \cos\theta < \cos\theta_*
\end{equation}
with 
\begin{equation}
\cos\theta_* = - 
\left(1-{3m\over\sqrt{r_*^2+a^2} }\right)\; \sqrt{1+{6m\over\sqrt{r_*^2+a^2}}},
\end{equation}
which in the range of interest can be more usefully written as 
\begin{equation}
\cos\theta_* = 
\left({3m\over\sqrt{r_*^2+a^2} }-1\right)\; \sqrt{1+{6m\over\sqrt{r_*^2+a^2}}}.
\end{equation}
This corresponds to selecting the union of  those up-going null geodesics that do not quite 
make it to  $r_\gamma^+$ and those  down-going null geodesics that do not quite make it to  $r_\gamma^-$.
The trapping cone solid angle is then
\begin{equation}
(\Delta \Omega_*)^{trapping}= 2\pi [\cos\theta_* - (-\cos\theta_*)] = 4\pi \; \cos\theta_*,
\end{equation}
and so we have the same result
\begin{equation}
(\Delta \Omega_*)^{trapping}= 4\pi \; \left({3m\over\sqrt{r_*^2+a^2} }-1\right)\; \sqrt{1+{6m\over\sqrt{r_*^2+a^2}}}.
\end{equation}

\paragraph{Emission from the other universe:}
For photon emission from  the other side of the wormhole ($r_* < r_\gamma^- = - \sqrt{(3m)^2-a^2}$), 
and considering emission to our asymptotic region $r\gg r_\gamma^+$ the photon would need to cross all three photon spheres. But we note that the minimum value of $H(r_\gamma)$, governing the narrowest escape cone, is
\begin{equation}
\min\{ H(r_\gamma)\} = \min\{H(r_\gamma^+),H(r_\gamma^0),H(r_\gamma^-)\} 
= \min\left\{27 m^2, {a^2\over 1-2m/a}, 27 m^2\right \} = 27 m^2.
\end{equation}
So for the escape cone we can then again say 
\begin{equation}
\Delta\Omega_* = 2\pi\left[ 1 -
 \left(1-{3m\over\sqrt{r_*^2+a^2} }\right)\; \sqrt{1+{6m\over\sqrt{r_*^2+a^2}}}\;\right].
\end{equation}
This boils down to saying that for photons coming in from the region exterior to that between the two unstable photon spheres we can safely neglect the region  between the two unstable photon spheres.
\enlargethispage{30pt}

Then for large negative $r_* \ll r_\gamma^-$ we have
\begin{equation}
\Delta\Omega_* =  {27\pi m^2\over r_*^2} \left(1-{2m\over|r_*|}\right) + \O(1/r_*^4).
\end{equation}
which completes the calculation.

\paragraph{Summary:} 
Note that it was the symmetry between the locations and optical areas of the two unstable photon spheres that allowed us to keep this explicit calculation relatively tractable. Note further that  the only null geodesics that  become trapped between the unstable photon spheres are a subset of  those that actually originate in that region.

\section{Conclusions}
What have we learnt from this discussion? Quite generally we have seen that in the presence of a single unstable photon sphere the escape cone solid angle for photons isotropically emitted from $r_*$ is quite generally given by 
\begin{equation}\Delta\Omega_* 
= 2\pi \left[ 1+\sign(r_*-r_\gamma) \sqrt{1-  {H(r_\gamma)\over H(r_*)}}\; \right],
\end{equation}
and in terms of the optical area
\begin{equation}\Delta\Omega_* 
= 2\pi \left[ 1+\sign(r_*-r_\gamma) \sqrt{1-  {\A(r_\gamma)\over \A(r_*)}}\; \right].
\end{equation}

For photons emitted at large distances from the central body we have the universal result
\begin{equation}
\Delta\Omega_* = 4\pi - {\sigma_{capture}\over r_*^2}\;(1-2m_\infty/r_*) + \O(1/r_*^4),
\end{equation}
where the capture cross section is
\begin{equation}
\sigma_{capture} = \pi \,H(r_\gamma)= {\A(r_\gamma)\over 4}.
\end{equation}

In contrast for near-horizon emission
\begin{equation}
\Delta\Omega_* =  {\sigma_{capture}  \over r_H^2} \;\exp[-\Phi(r_H)] \;\kappa_H \; [r_*-r_H] + \O([r_*-r_H]^2),
\end{equation}
which explicitly depends on the surface gravity. 

If one prefers to work with escape probabilities then 
\begin{equation}
P_* = 1 - {\sigma_{capture}\over 4\pi r_*^2}\;(1-2m_\infty/r_*) + \O(1/r_*^4),
\end{equation}
and
\begin{equation}
P_* =  {\sigma_{capture}  \over 4\pi r_H^2} \;\exp[-\Phi(r_H)] \;\kappa_H \; [r_*-r_H] + \O([r_*-r_H]^2).
\end{equation}
These results are general purpose tools whose only real restriction is the assumption of static spherical symmetry and the presence of a single photon sphere.
For multiple photon spheres we have provided some general guidance, and have have worked out one example (the black bounce spacetime) in some detail.
There are potentially many other examples that could be worked out in detail.

Of course the restriction to static spherical symmetry is exactly what allowed us to avoid resorting to numerics and to keep good analytic control over the situation.
In the long run one would really wish to address Kerr and Kerr-like spacetimes~\cite{Kerr:1963, Kerr:1965, Kerr:book, Kerr:intro, Kerr:2007, Kerr:ansatz}.
Doing so is much trickier, see~\cite{Perlick:2021, Perlick:2010, Tsupko:2017, Grenzebach:2014} amd~\cite{KdS:escape, KN:escape, Kerr:near-extremal, Zulianello:2020, Amo:2023}. Current efforts rather quickly resort to numerics.

\bigskip
\hrule\hrule\hrule


\section*{Acknowledgements}

JB was supported by a Victoria University of Wellington PhD Doctoral Scholarship.
\\
MV was directly supported by the Marsden Fund, 
via a grant administered by the Royal Society of New Zealand.

\bigskip
\hrule\hrule\hrule
\clearpage
\addtocontents{toc}{\bigskip\hrule}

\null
\vspace{-50pt}
\setcounter{secnumdepth}{0}
\section[\hspace{14pt}  References]{}
%

\end{document}